\documentclass[preprint,prl,showpacs,preprintnumbers]{revtex4}
\usepackage{amsfonts}
\usepackage{graphicx}
\usepackage{dcolumn}
\usepackage{bm}
\usepackage{amsmath}
\usepackage{multirow}

\begin{document}

\preprint{APS/123-QED}
\title{Mode-balancing far field control of light localization in nanoantennas}
\author{Alexis Devilez}
\author{Brian Stout}
\author{Nicolas Bonod}
\email{nicolas.bonod@fresnel.fr}
\affiliation{Institut Fresnel, Aix-Marseille Universit\'e, CNRS, Domaine universitaire de
Saint-J\'er\^ome, 13397 Marseille, France\\
}
\date{\today}

\begin{abstract}
Light localization is controlled at a scale of $\lambda$/10 in the harmonic regime from the far field domain in a plasmonic nanoantenna. The nanoantenna under study consists of 3 aligned spheres 50 $nm$ in diameter separated by a distance of 5 $nm$. By simply tuning the orientation of an incident plane wave, symmetric and antisymmetric mode-balancing induces a strong enhancement of the near field intensity in one cavity while nullifying the light intensity in the other cavity. Furthermore, it is demonstrated that the dipolar moment of a plasmonic particle can be fully extinguished when strongly coupled with a dimer of identical nanoparticles. Consequently, optical transparency can be achieved in an ultra-compact symmetric metallic structure.
\end{abstract}

\pacs{Valid PACS appear here}
\maketitle



Metallic nanostructures offer the opportunity to strongly focus light and
to enhance light-matter interactions at the nanometer scale via the excitation
of localized surface plasmon resonances. In 2002, Stockman et al.
demonstrated the dynamic control of the field enhancement at
nanometer scale in arbitrary nanostructures \cite{stocloc} by adjusting the phase and the
polarization of an excitation pulse. Active control of field localization has been experimentally
demonstrated through the use of an iterative learning algorithm to shape the
femtosecond excitation pulse \cite{BrixNat2007}. Deterministic structures such as arrays of
spherical nanoparticles have been investigated by Koenderinck et al. \cite{lithoarray} to serve as
unique and reproducible lithographic masks for imprinting different near field
patterns. They varied the angle of incidence and polarization of the
continuous excitation beam in order to produce a variety of patterns in a
photosensitive substrate. These studies convincingly demonstrated that incident
beam shaping can induce different field distributions in a single metallic
nanostructure.

In 2006, Le Perchec et al. controlled the localization of
light in coupled slits milled in a metallic substrate \cite{Barbaraswitch}. The slits had a
thickness of 200 $nm$ and were separated by a distance of 500 $nm$. When
illuminated in oblique incidence in the infrared spectrum, they coupled
incident propagating light to both symmetric and anti-symmetric modes.
By the use of a second identical frequency propagating wave in opposite
incidence, they could switch the near field enhancement from one slit to the
other via phase adjustments. This simple system provided a means to
conceive subwavelength optical switch components controlled from the far
field region. Very recently, Volpe et al. \cite{quidantswitch} investigated spatial phase
modulations of high order beams such as Laguerre-Gaussian beams in order to
control the near field in plasmonic nanostructures. They considered two 50
$nm$ gaps formed by three 500 $nm$ aligned gold bars and demonstrated that
spatial phase modulation of the incident beam enables switching light on and
off in either of the two sites. The aim of this paper is to theoretically study a linear
trimer of identical spherical nanoparticles to tune the localization of
light at the nanometer scale ($\sim $55 $nm$). In particular, we investigate
the near field enhancement occuring between gold nanospheres 50 $nm$ in diameter.
The three spherical nanoparticles are arranged to form two identical
cavities in which the near field can be strongly enhanced. This provides a
simple system with two coupled nanocavities separated by a distance of about
a tenth of the incident wavelength. It will be shown that this a priori
symmetric system enables the concentration of light in a single cavity while employing a single illumination plane wave.

\begin{figure}[htbp]
\includegraphics[width=9cm]{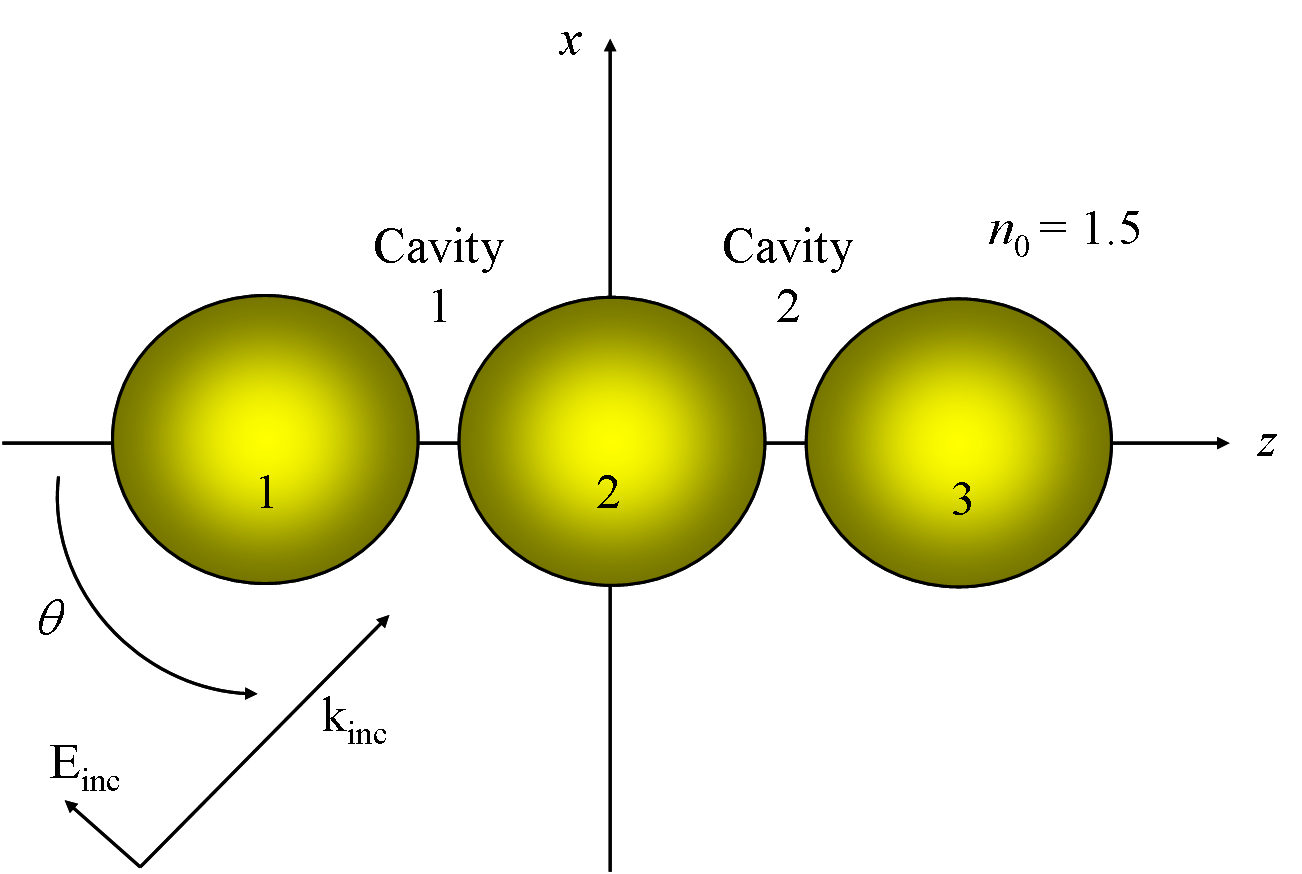}
\caption{Layout of the numerical experiment. Three identical gold nanospheres 50 $nm$ in diameter, embedded in a dielectric background of refractive index $n_0$ = 1.5, are placed along the $z$-axis. 
The system is illuminated by a plane wave with a TM polarization with an angle of incidence $\theta$ between the propagation direction and the $z$-axis. Calculations are performed using generalized Mie theory combined with an analytic multiple scattering formulation.}
\label{fig:schema}
\end{figure}

We investigate a system consisting of three identical gold nanospheres placed along the $x$-axis with a diameter of 50 $nm$ and embedded in a polymer substrate with a refractive index of 1.5 (Fig.~\ref{fig:schema}). The choice of the structure is motivated by the study of Bidault et al. \cite{bidaultADN08} where well-defined groupings of Au nanoparticles with controlled nanometer spacing were synthesized by hybridizing monoconjugated DNA in a polymeric material of refractive index $n_0$ = 1.5. This bottom-up technology opens up a way to conceive coupled cavities made of nanoparticles separated by a few nanometers, currently unobtainable via classical lithography techniques. 

This model system is illuminated by a plane wave with TM polarization (the magnetic field is perpendicular to the plane presented in Fig.~\ref{fig:schema}) and an angle of incidence,  $\theta$,  tunable between 0 (propagation along the $z$-axis) and 90 degrees. 
The strong contrasts of near field intensities at scales much smaller than the incident wavelength require a full electromagnetic study based on solving the Maxwell equations. Rigorous Lorentz-Mie theory is employed in order to accurately calculate the near field enhancement in the vicinity of a single spherical metallic nanoparticle. With this technique, electromagnetic fields inside the sphere and in the surrounding medium are expanded using basis sets of Vector Spherical Harmonics (VSHs), which allow an analytic satisfaction of the boundaries conditions. 
The expansion of an arbitrary vector field is given in Eq. (\ref{sol hetero}) in terms of the Vector Spherical Harmonics. The VSHs provide an angular description of the vectorial field and are defined  in terms of the Scalar Spherical Harmonics $Y_{n,m}(\theta,\phi)$ in Eq. (\ref{harm_sph_vect}). The $E^{(X,Y,Z)}_{n,m}$ are their respective coefficients that describe the radial dependency of the field according to the medium properties. Interactions between particles are taken into account by invoking the translation/addition theorem in order to solve the Foldy-Lax multiple scattering equations \cite{FoldyLax,Stout08}. The truncation of the basis sets, necessary for a numerical solving of this system of equations, is determined from the size parameter of the spheres and the strength of electromagnetic couplings between the particles, and results in a finite dimensional T-matrix describing the electromagnetic response of the entire system \cite{wiscombe}. In this study, high multipolar orders are required to accurately reconstruct the field which will be taken into account by imposing $n_{max}=10$ in Eq. (\ref{sol hetero}). The dispersion of the dielectric constant of gold is interpolated from the Palik database \cite{palik}.

\begin{widetext}
\begin{equation} 
\overrightarrow{\mathbf{E}}(r) =\sum_{n=0}^{\infty }\sum_{m=-n}^{n}\left[
E_{n,m}^{(Y)}(r)\overrightarrow{\mathbf{Y}}_{n,m}(\theta ,\phi )+E_{n,m}^{(X)}(r)\overrightarrow{\mathbf{X}}_{n,m}(\theta ,\phi )+E_{n,m}^{(Z)}(r)\overrightarrow{\mathbf{Z}}_{n,m}(\theta ,\phi )\right] 
\label{sol hetero}
\end{equation}
\end{widetext}

\begin{subequations}
\begin{equation}
\overrightarrow{\mathbf{Y}}_{n,m}(\theta,\phi) \equiv Y_{n,m}(\theta ,\phi)%
\widehat{\mathbf{r}} \\
\label{suba}
\end{equation}
\begin{equation}
\overrightarrow{\mathbf{X}}_{n,m}(\theta,\phi)  \equiv\overrightarrow {%
\mathbf{Z}}_{n,m}(\theta,\phi)\times\widehat{\mathbf{r}} \\
\label{subb}
\end{equation}
\begin{equation}
\overrightarrow{\mathbf{Z}}_{n,m}(\theta,\phi)  \equiv\frac {r%
\overrightarrow{\nabla}Y_{n,m}(\theta,\phi)}{\sqrt{n(n+1)}}
\label{subc}
\end{equation}
\label{harm_sph_vect}
\end{subequations}

\begin{figure}[htbp]  
\includegraphics[width=8cm]{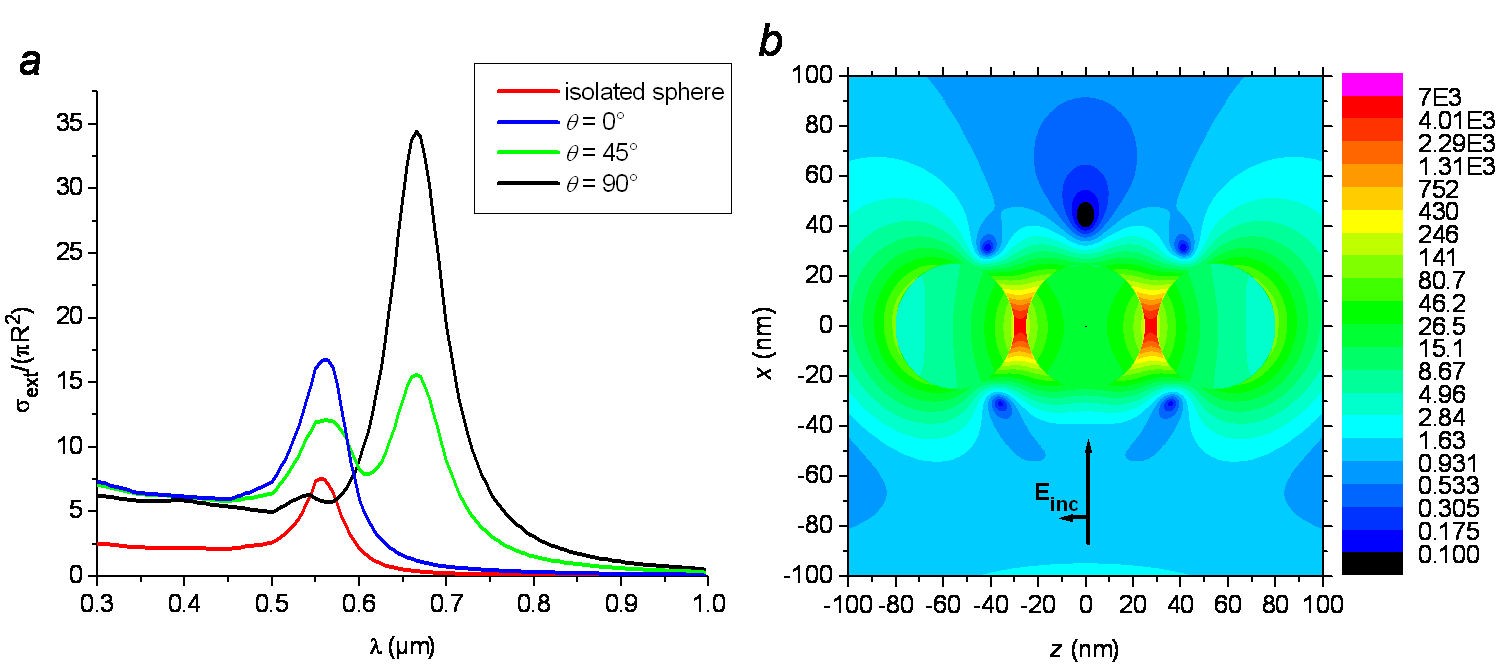}
\caption{(a) Extinction cross-section of a chain of three gold particles of diameter $D$ = 50 $nm$ (red line) separated by $w$ = 5 $nm$ in the full electromagnetic approach. Different angles of incidence are displayed: $\theta = 0^{\circ}$ (blue line), $\theta = 45^{\circ}$ (green line) and $\theta = 90^{\circ}$ (black line). The isolated sphere extinction cross-section is also displayed (red line). (b) Electric field intensity $\|E\|^2 /\|E_{inc}\|^2$ map in log scale for a trimer illuminated at $\lambda$ = 665 $nm$ under normal incidence}
\label{fig:chploin}
\end{figure}

We first investigate the far field response of the trimer by rigorously calculating the spectral response of the extinction cross-section for different angles of incidence (Fig.~\ref{fig:chploin}$a$). For an incident angle $\theta = 0^{\circ}$ (blue line) corresponding to an illumination along the $z$-axis, only one maximum appears at the wavelength of 560 $nm$, which is almost the wavelength of resonance of the isolated sphere (red line). Assuming an incident electric field along the $x$-axis, it can be deduced that this resonance is due to in-phase plasmon oscillations occuring transversally to the chain axis resulting in weak couplings between the individual nanoparticles resonances. At normal incidence ($\theta = 90^{\circ}$: black line), a stronger resonance occurs at the wavelength of 665 $nm$, corresponding to the well-known red-shifted coupled resonance of in-phase longitudinal modes \cite{prop_gold_nnpt}. Fig.~\ref{fig:chploin}$b$ illustrates the strong near field enhancement between the spheres when the trimer is illuminated at $\lambda$ = 665 $nm$. When the chain is illuminated at normal incidence, only in-phase longitudinal oscillations of the polarization moments are excited. These oscillations induce surface charges of opposite sign in the dielectric inter-particle gap responsible for the strong electric field enhancement. Such longitudinal modes have been studied extensively in the context of dimers of metallic spheres and nanoshells \cite{Bergman_th, prop_gold_nnpt,Ag_nnpt_pairs,stocknanolens,peanuts08,HibridNorland04}. We remark that for the case of normal incidence, there exists a weak second peak at higher frequency which is made possible thanks to quadrupolar response contributions. At intermediate angles of incidence (like the case $\theta = 45^{\circ}$ displayed in Fig.~\ref{fig:chploin}$a$), we observe two peaks at frequencies close to those observed respectively at $\theta = 0^{\circ}$ and $\theta = 90^{\circ}$.

Table.~\ref{tab:table} illustrates the dipolar eigenmodes that may be excited in a linear trimer. 
Although the dipolar approximation fails to describe accurately the resonance when the particles are strongly coupled, it enables a clear distinction between modes that are perpendicular (transverse) and parallel (longitudinal) to the chain axis \cite{aggrscat93}. 
From the extinction cross-section displayed in Fig.~\ref{fig:chploin}$a$, one can observe that only the in-phase modes denoted T1 and L1 are clearly visible. This observation may seem surprising since except for illumination at $\theta = 0^{\circ}$ or $\theta = 90^{\circ}$ where the symmetries of the configuration impose in-phase modes, coupling between the incident field and opposite-phase modes may be possible when illuminating the system at oblique incidence. 
Inspection of the opposite-phase modes in Table.~\ref{tab:table} shows however that the total dipolar moment involved in these modes is weaker than that occurring for in-phase modes. Consequently, these modes are not easily coupled to the far field and cannot be observed in extinction cross-section plots.

\begin{table}[htbp]
\caption{\label{tab:table}Possible dipolar eigen modes for a linear chain of three plasmonic particles }%

\begin{tabular}  {|l|cccc|c|}
\hline
\multirow{3} {*} {Transverse eigenmodes}  &  T1  & $  \uparrow$  &  $  \uparrow$  &  $  \uparrow$  &    In-phase mode\\
\cline{2-6}
                                      		&    T2 &  $\downarrow$  &  $\uparrow$  &  $\uparrow$ &  Opposite-phase \\
                                      		 &   T3 & $\uparrow$  &  $\downarrow$  &  $\uparrow$   &   modes                    \\
\hline

\multirow{3} {*} {Longitudinal eigenmodes} & L1 &  $\rightarrow$ &   $\rightarrow$   & $\rightarrow$  &   In-phase mode\\
\cline{2-6} 
 					&   L2 & $\leftarrow$  &  $\rightarrow$  &  $\rightarrow$ & Opposite-phase \\
                                      		 &   L3  & $\leftarrow$  &  $\rightarrow$ &  $\leftarrow$   &   modes                    \\

\hline
\end{tabular}

\end{table}

\begin{figure}[htbp]  
\includegraphics[width=8.5cm]{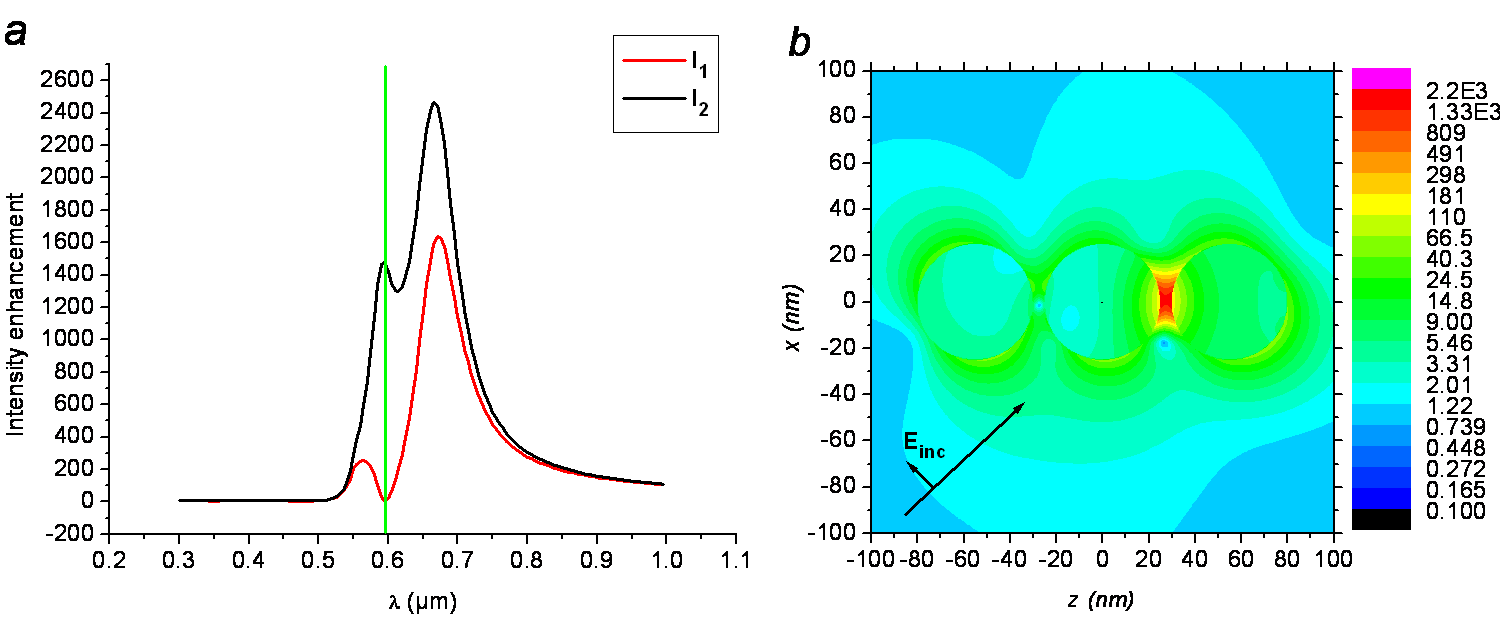}
\caption{(a) Intensity enhancement in cavity 1 (red line) and cavity 2 (black line) for $\theta = 45^{\circ}$. The vertical line indicates the wavelength of 595 $nm$ where the intensity contrast is maximum. (b) Electric field intensity, $\|E\|^2/\|E_{inc}\|^2$ ,  map in log scale at $\lambda$ = 595 nm and $\theta = 45^{\circ}$.}
\label{fig:chploc}
\end{figure}

Let us emphasize that we are interested in the vanishing of the near field enhancements in one of the cavities. This purpose requires an excitation via the far field of longitudinal opposite-phase modes, modes that are associated with a weak dipolar moment. Fig.~\ref{fig:chploc}$a$ displays the near field intensity enhancement calculated at the center of the cavities as a function of the incident wavelength, when incident angle $\theta$ is equal to 45 degrees. These plots show that the field intensities in the two cavities are quite different, but most interestingly they reveal the counterintiutive occurence that at $\lambda$ = 595 $nm$ (green line) a rather strong field intensity enhancement in gap 2 is concurrent with a near vanishing of the field in gap 1 (ratio larger than 500). This phenomenon is even more conspicuous in the near field intensity map displayed in Fig.~\ref{fig:chploc}$b$ in logarithmic scale that clearly shows that at $\lambda$ = 595 $nm$, the light intensity is confined in only one site. This is the most striking result of this study: the sub-wavelength cavity 1 is turned-off and the light intensity is almost entirely localized inside cavity 2 while both cavities are separated by only 55 $nm$. It appears that this phenomenon cannot be observed from the far field response since the extinction cross section (green line in Fig.~\ref{fig:chploin}$a$) does not exhibit any notable feature. We next move towards the full understanding of the longitudinal eigen modes involved in this phenomenon, which will require the extraction of longitudinal terms in the multipole formalism.

The extraction of the longitudinal components of the field requires that we focus attention on the Scalar Spherical Harmonics that are defined in Eq.  (\ref{scal_harm_sph}) in terms of the associated Legendre polynomials $P_{n}^{m}(\cos\theta)$.

\begin{equation}
Y_{nm}(\theta,\phi)   \equiv  c_{n,m}P_{n}^{m}\left(  \cos\theta\right)  \exp(im\phi)
\label{scal_harm_sph}
\end{equation}
where
\begin{equation}
 c_{n,m} \equiv   \left[  \frac{2n+1}{4\pi}\frac{(n-m)!}{(n+m)!}\right]  ^{\frac{1}{2}}
\label{eq:normalization}
\end{equation}

The associated Legendre polynomials $P_{n}^{m}(\cos\theta)$  have the following property:
\begin{subequations}
\begin{equation}
\forall m\neq 0\text{ \ \ \ }P_{n}^{m}(\pm 1)=0
\end{equation}
\begin{equation}
P_{n}^{0}(\pm 1)=\left( \pm 1\right) ^{n}
\end{equation}
\label{legendre}
\end{subequations}

\begin{figure}[htbp]  
\includegraphics[width=8.5cm]{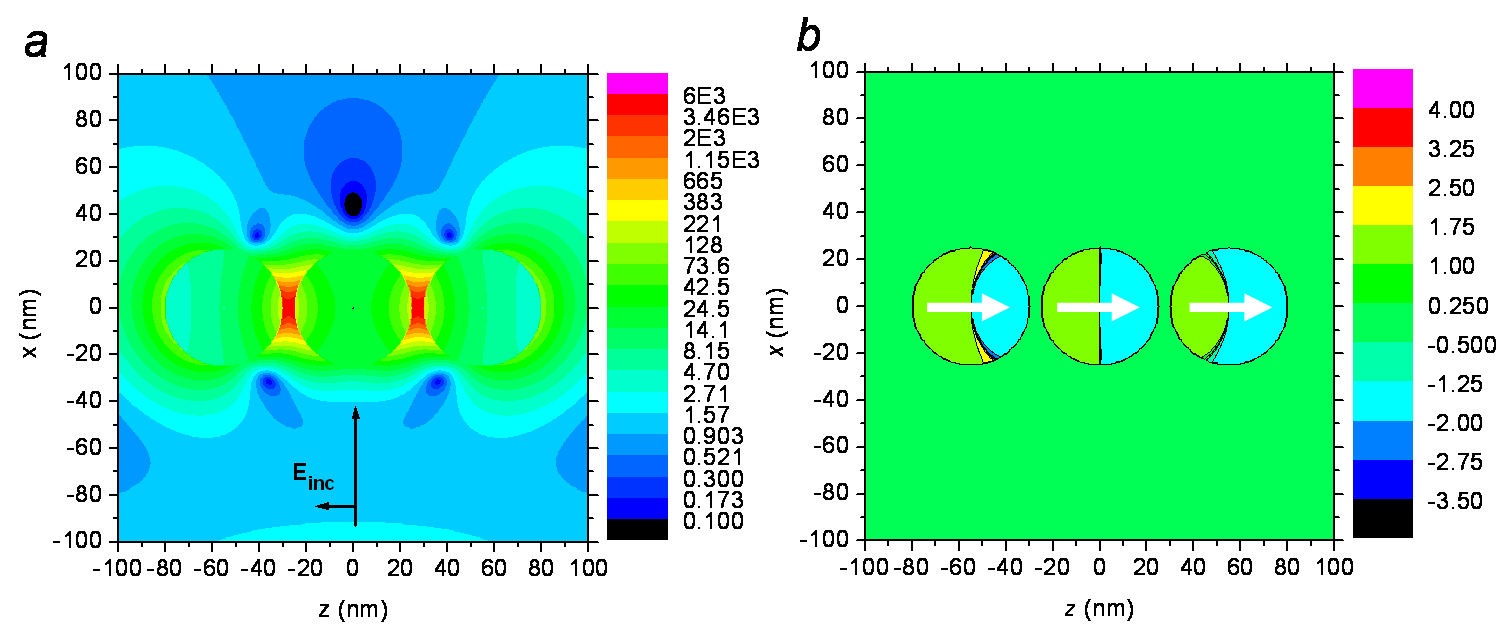}
\caption{(a) Electric field intensity $\|E\|^2 /\|E_{inc}\|^2$ map in log scale of the longitudinal contributions reconstructed imposing $(n,m) = (n,0)$ in Eq. (\ref{sol hetero}) and (b) phase of the field reconstructed with the terms $E_{n,0}^{(Y)}(r)Y_{n,0}$ in Eq. (\ref{sol hetero}) at $\lambda$ = 665 $nm$ and $\theta = 90^{\circ}$.}
\label{fig:incnorm}
\end{figure}

Eq. \ref{legendre} demonstrates that the field distribution along the $z$-axis, corresponding to $\cos\theta = \pm 1$ can be entirely described by taking the scalar spherical harmonics $Y_{n,0}$. Fig.~\ref{fig:incnorm}$a$ shows the field reconstructed by imposing $m = 0$ in Eq.(\ref{sol hetero}) with the same conditions of illumination as in Fig.~\ref{fig:chploin}$b$. Comparison of these two maps illustrates that the longitudinal in-phase resonances can be fully described when imposing $m=0$ and that this method can be considered as a relevant way to study the longitudinal modes of the structure. It must be stressed that these longitudinal contributions include the three directions of the field $\overrightarrow{\mathbf{X}}_{n,0}$, $\overrightarrow{\mathbf{Y}}_{n,0}$ and $\overrightarrow{\mathbf{Z}}_{n,0}$. The contribution of the electric field oriented along the chain axis is then given by the terms $E_{n,0}^{(Y)}(r)Y_{n,0}$ in Eq.(\ref{sol hetero}). Fig.~\ref{fig:incnorm}$b$ shows the phase of the field reconstructed with the terms $E_{n,0}^{(Y)}(r)Y_{n,0}$. The in-phase oscillations of the longitudinal mode in the structure are clearly highlighted. Let us remark that the polarization moment inside the particles shows a clear dipolar-like behaviour.

\begin{figure}[htbp]  
\includegraphics[width=8.5cm]{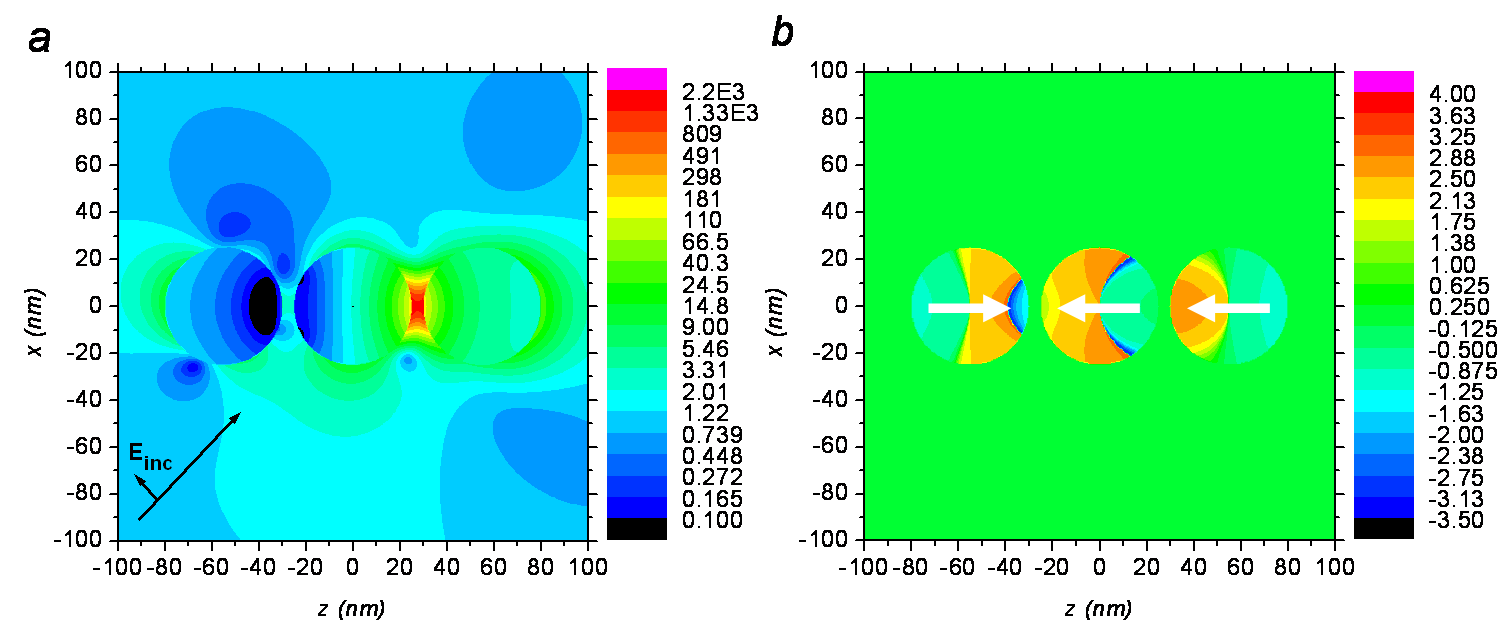}
\caption{$(a)$ Electric field intensity $\|E\|^2 /\|E_{inc}\|^2$ map in log scale of the longitudinal contributions reconstructed imposing $(n,m) = (n,0)$ in Eq. (\ref{sol hetero}) and $(b)$ phase of the field reconstructed with the terms $E_{n,0}^{(Y)}(r)Y_{n,0}$ in Eq. (\ref{sol hetero}) at $\lambda$ = 565 $nm$ and $\theta = 45^{\circ}$.}
\label{fig:incobl}
\end{figure}

We now apply this formalism to study the phenomenon observed in Fig.~\ref{fig:chploc}. Fig.~\ref{fig:incobl}$a$ displays the longitudinal multipole contributions to the field at $\lambda$ = 595 $nm$ and $\theta = 45^{\circ}$ obtained by reconstructing the field intensity map with $(n,m)=(n,0)$. It highlights that the vanishing of the near field in cavity 1 associated with the high intensity enhancement in cavity 2, which are the striking features observed in Fig.~\ref{fig:incobl}$a$, are essentially obtained. It shows that the intensity contrast observed is due to the excitation of longitudinal modes in the structure. 
More precisely, the phase of the longitudinal electric field displayed in Fig.~\ref{fig:incobl}$b$, obtained under the same conditions as Fig.~\ref{fig:incnorm}$a$, shows that the mode L2 depicted in Table.~\ref{tab:table} is predominantly excited. The oblique illumination has enabled the coupling of the incident light to an opposite-phase mode that is responsible for a vanishing of the field in cavity 1 while the field in cavity 2 is strongly enhanced. Let us emphasize that the electric field oscillations are principally dipolar in nature and that the longitudinal mode observed in Fig.~\ref{fig:chploc} involves predominantly dipolar contributions. 
Unlike the case illustrated in Fig.~\ref{fig:incnorm}$a$, Fig.~\ref{fig:incobl}$a$ clearly shows a weak longitudinal induced polarization inside sphere 1. The asymmetry of the electric field amplitudes inside the spheres in Fig.~\ref{fig:incobl}$a$ indicates not only that the L2 mode is excited, but that a combination of the three longitudinal modes (Table.~\ref{tab:table}) is involved. This result shows that it is possible to nearly extinguish the longitudinal moment of a metallic nanoparticle by simply coupling this particle with a dimer of identical particles. It means that optical transparency \cite{PlEITmeta} can be achieved with particles which do not support transverse modes, i.e. ellipsoidal or rectangular nanoparticles. Consequently, this study demonstrates that the dipolar moment of a plasmonic particle can be fully extinguished in ultra compact symmetric structures.

This study demonstrates that antisymmetric modes can be excited in a linear trimer of identical particles by simply tuning the angle of incidence. Furthermore, we show that in a narrow frequency range, symmetric and antisymmetric mode-balancing extinguishes the dipolar longitudinal moment of a metallic nanosphere coupled to a dimer composed of identical nanospheres. At this frequency, the total longitudinal moment of the trimer is weak and this phenomenon is not associated with an enhancement of the scattering cross-section of the trimer. This nearly zero longitudinal moment of sphere 1 leads to a near vanishing light intensity in cavity 1, and the high longitudinal moment of sphere 3 leads to strong enhancement of light intensity in cavity 2. Finally, we wish to emphasize that this phenomenon occurs in a simple fully sub-wavelength system by employing straighforward far-field control in the harmonic regime. Specifically, it enables the control of light localization between two identical nanosites separated by a distance of 55 $nm$, leading to a spatial resolution of $\lambda$/10.

\begin{acknowledgments}
The authors thank S. Bidault and R. C. McPhedran for stimulating discussions.
\end{acknowledgments}

\bibliography{switch}

\end{document}